 \documentclass[smallabstract,smallcaptions]{dccpaper}

\usepackage{epsfig}
\usepackage{citesort}
\usepackage{amsmath}
\usepackage{amssymb}
\usepackage{color}
\usepackage{url}
\usepackage{booktabs}
\usepackage{multirow}
\usepackage{makecell}
 \usepackage{threeparttable}

\newlength{\figurewidth}
\newlength{\smallfigurewidth}

\begin{document}

\title
{\large
\textbf{Perceptual-oriented Learned Image Compression with Dynamic Kernel}
}



\author{
Nianxiang Fu$^{1}$, Junxi Zhang$^{2}$, Huairui Wang$^{2}$ and Zhenzhong Chen$^{1, 2 \ast}$\thanks{$^{\ast}$Corresponding author.}\\[0.5em]
{\small\begin{minipage}{\linewidth}\begin{center}
$^{1}$Hubei Luojia Laboratory, Wuhan University,  Wuhan, China \\
$^{2}$School of Remote Sensing and Information Engineering, Wuhan University, Wuhan, China\\
\url{zzchen@ieee.org} 
\end{center}\end{minipage}}
}
\maketitle
\thispagestyle{empty}

\begin{abstract}
In this paper, we extend our prior research named DKIC and propose the perceptual-oriented learned image compression method, PO-DKIC. Specifically, DKIC adopts a dynamic kernel-based dynamic residual block group to enhance the transform coding and an asymmetric space-channel context entropy model to facilitate the estimation of gaussian parameters. Based on DKIC, PO-DKIC introduces PatchGAN and LPIPS loss to enhance visual quality. Furthermore, to maximize the overall perceptual quality under a rate constraint, we formulate this challenge into a constrained programming problem and use the Linear Integer Programming method for resolution. The experiments demonstrate that our proposed method can generate realistic images with richer textures and finer details when compared to state-of-the-art image compression techniques.
\end{abstract}

\section{Introduction}

Nowadays, multimedia data is growing exponentially, rending a contradiction between a huge volume of data and the limited bandwidth. Image compression plays a critical role in image storage and transmission. After decades of development, some traditional compression approaches have achieved promising compression performance, such as JPEG \cite{wallace1992jpeg}, BPG \cite{bellard2017bpg}, and VVC \cite{chen2019algorithm}. These methods are typically designed in a hybrid style and rely on human-crafted techniques, such as transform, quantization, and entropy coding. Due to the dependency between these modules, the coding efficiency of traditional methods is limited by the separately optimized framework and the need for more flexibility to adapt to different image contents. 
In recent years, there has been increasing interest in applying deep learning techniques to image compression. Different from traditional approaches, the leaned compression framework can be jointly optimized in an end-to-end manner.
Moreover, learning-based codecs can easily be optimized according to perception-related metrics to generate more realistic images. Some recent learning-based image compression approaches have outperformed VVC in terms of Peak Signal-to-Noise Ratio (PSNR) and Multi Scale-Structural Similarity Index Measure (MS-SSIM). The learned compression shows strong potential to develop the next-generation image compression framework. 

In the traditional codec, transform coding decorrelates the image signal by well-designed algorithms, such as Discrete Cosine Transform (DCT), Discrete Sine Transform (DST), and so on. The learning-based compression utilizes the learned transform coding to map the image to compact latent representations. CNN-based methods tend to stack convolutional layers to aggregate information in a fixed range according to kernel size. Our prior work DKIC \cite{wang2023dynamic} introduced Lite DCN (LDCN) to break the limitation of fixed range spatial aggregation. Moreover, we explore the energy response and the spatial correlation in the latent and define an asymmetric spatial-channel entropy model to remove statistical redundancy efficiently. Experimental results demonstrate that DKIC achieves superior rate-distortion performance on three benchmarks compared to the state-of-the-art learning-based methods. However, the optimization distortion of DKIC is mean square error, which may cause a mismatch between the subjective and objective perceptions. In this paper, we enhance DKIC along with the Generative Adversarial Networks (GAN) \cite{wang2018esrgan} loss and LPIPS loss to generate more realistic images.

\section{Method}
\subsection{Architecture}
\begin{figure*}[t]
    \centering
    \includegraphics[width=\textwidth]{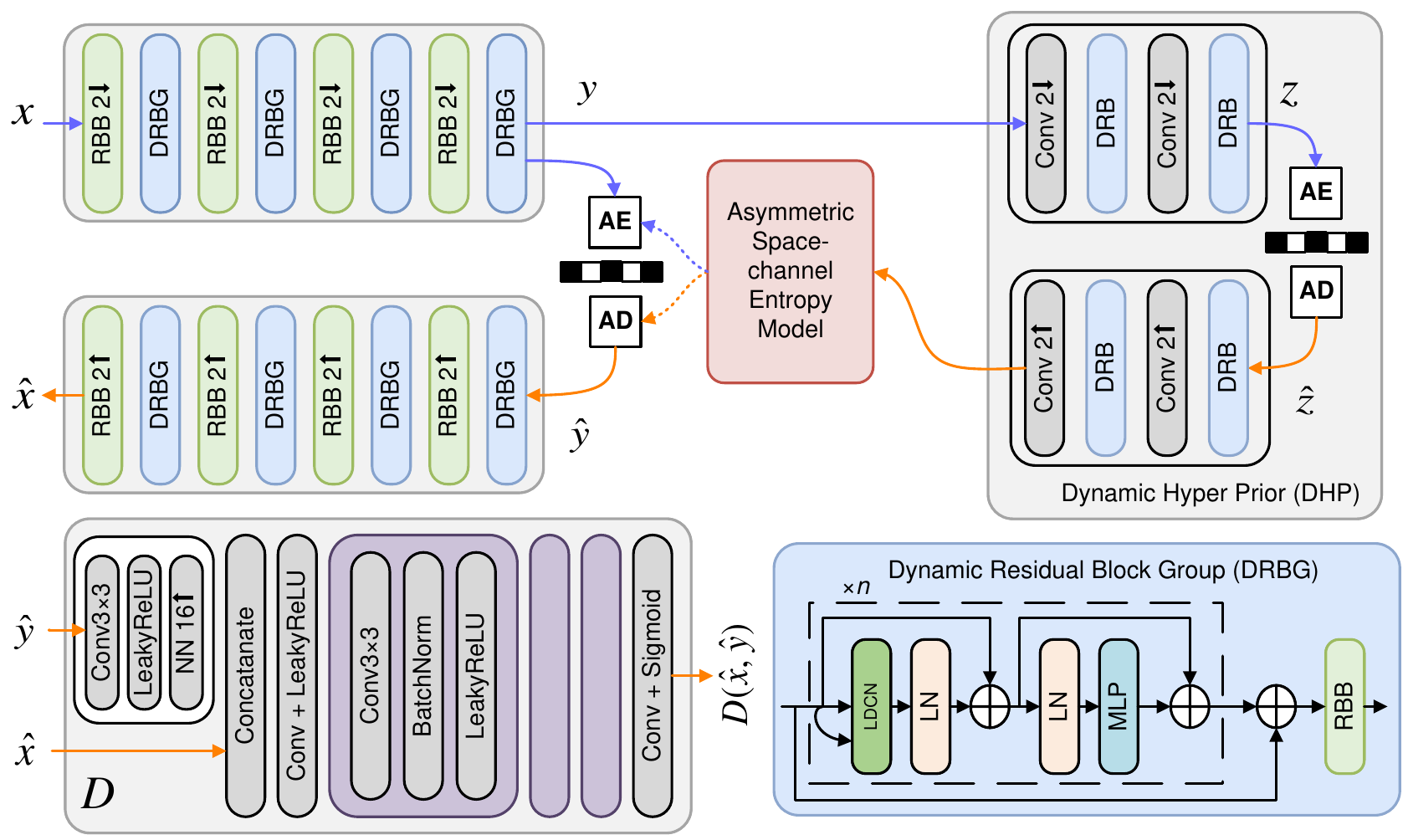}
    \caption{The framework of our image compression. The upper part is DKIC. The lower part shows the adversarial training.} 
    \label{fig:dkic}
\end{figure*}

The architecture is based on our previous work DKIC \cite{wang2023dynamic}, and the overall compression framework is shown in Figure \ref{fig:dkic}.

DKIC adopts a dynamic kernel-based Dynamic Residual Block Group (DRBG) with dynamic spatial aggregation capacity, which generates kernel offsets and modulated kernel weights to capture relevant information within content-dependent ranges, improving the transform coding. Furthermore, to facilitate the estimation of Gaussian parameters, DKIC adopts an Asymmetric Space-Channel Context Entropy Model, which is a generalized coarse-to-fine entropy model that considers global context, channel-wise information, and spatial context in a coarse-to-fine manner.

In addition, in order to generate more realistic images, following HiFiC \cite{mentzer2020high}, we treat the decoder as a conditional generator, coupled with a conditional discriminator, and we use the same conditional discriminator structure as HiFiC. We employ adversarial training to optimize the entire framework.

\subsection{Perceptual Loss Function}
We adopt the rate-constrained Rate-Distortion (RD) optimization framework from HiFiC:
\begin{equation}\label{eq:1}
\begin{aligned}
\mathcal{L}=\mathcal{L}_{\mathcal{D}}+\lambda(\mathcal{R}, \mathcal{R}^t)\cdot \mathcal{R},
\end{aligned}
\end{equation}
where $D$ and $R$ represent the perceptual distortion and rate terms, respectively. The multiplexer function $\lambda(\mathcal{R},\mathcal{R}^t)$ is conditioned on the specified target bitrate $\mathcal{R}^t$, and it can be expressed as:
\begin{equation}\label{eq:2}
\begin{aligned}
\lambda(\mathcal{R},\mathcal{R}^*)=\begin{cases}\lambda_\alpha,&\mathcal{R}\geq\mathcal{R}^*\\\lambda_\beta,&\mathcal{R}<\mathcal{R}^*\end{cases},
\end{aligned}
\end{equation}
where $\mathcal{R}^*$ denotes the target bitrate. Our perceptual distortion loss function can be summarized as:
\begin{equation}\label{eq:3}
\begin{aligned}
\mathcal{L}_{\mathcal{D}}=w_1\times\mathcal{L}_{rec}+w_2\times\mathcal{L}_{lpips}+w_3\times\mathcal{L}_{gan},
\end{aligned}
\end{equation}
where the perceptual loss $\mathcal{L}_{lpips}$ is LPIPS loss.
$\mathcal{L}_{rec}$ is MSE loss. $\mathcal{L}_{gan}$ is the GAN loss. $w_i$ is the weight of each loss.

As for the GAN loss, referring to our previous work \cite{wang2022perceptual}, we follow PatchGAN \cite{ma2020structure} to adopt Markovian discriminator to focus on modeling high-frequencies at the scale of patches and utilize Relativistic GAN loss \cite{wang2018esrgan} which helps to learn sharp edges and more detailed textures. We denote the Patch-wise Relativistic average Discriminator as $D_{Pat}$, which can be formulated as:
\begin{equation}\label{eq:4}
\begin{aligned}
D_{Pat}(x_r,x_f)=\sigma(C_{Pat}(x_r-\mathbb{E}_{x_f}[C_{Pat}(x_f)])),
\end{aligned}
\end{equation}
where $\sigma$ is the sigmoid function, $C_{Pat}(X)$ is the patch-wise discriminator output, $x_f$ and $x_R$ are the output images of the DKIC and the ground truth, respectively. The GAN loss function can be stated as follows:
\begin{equation}\label{eq:5}
\begin{aligned}
&\mathcal{L}_{D}= -\mathbb{E}_{x_r}[log(D_{Pat}(x_r,x_f))]-\mathbb{E}_{x_f}[log(1-D_{Pat}(x_f,x_r))], \\
&\mathcal{L}_{G}= -\mathbb{E}_{x_{r}}[1-log(D_{Pat}(x_{r},x_{f}))]-\mathbb{E}_{x_f}[log(D_{Pat}(x_f,x_r))].
\end{aligned}
\end{equation}

\subsection{Resource Allocation}
The task of this challenge is to determine the allocation rate for each compressed image to maximize the overall perceptual quality while adhering to a predefined rate constraint. Achieving this constraint with a model trained using a single $\lambda$ can be challenging when dealing with diverse datasets. To address this issue, we trained multiple models with different quality levels using various $\lambda$ values. With the objective of constrained optimization, we have transformed this task into a constrained programming problem:
\begin{equation}\label{eq:6}
\begin{aligned}
\begin{gathered}\arg\min\sum_{i=1}^N\sum_{j=1}^MLPIPS_i(Q_j)\times x_{ij}\\
s.t.\quad\frac{1}{N}\sum_{i=1}^N\sum_{j=1}^MR_i(Q_j)\times x_{ij}\leq T\\
\forall i\sum_{j=1}^Mx_{ij}=1\end{gathered},
\end{aligned}
\end{equation}
where $Q_j$ denotes the quality level of the compression model trained with $\lambda_j$. $LPIPS_i(Q_j)$ and $R_i(Q_j)$ are the corresponding lpips and rate cost. $x_{ij}$ represents the flag whether $j_{th}$ quality  is chosen for compressing the $i_{th}$ image. $T$ denotes the total target average bpp. $N$ denotes the number of images. $M$ denotes the number of quality levels.

\begin{table}[htbp]
\renewcommand{\arraystretch}{1.5}
\caption{\centering Objective results comparison on CLIC 2024 Validation set. Act BPP denotes actually bpp.}
\label{tab:obj-results-table}
\resizebox{\textwidth}{!}{%
\begin{tabular}{|c|cccc|cccc|cccc|}
\hline
\multirow{2}{*}{Method Name} &
  \multicolumn{4}{c|}{0.075 bpp} &
  \multicolumn{4}{c|}{0.15 bpp} &
  \multicolumn{4}{c|}{0.3 bpp} \\ \cline{2-13} 
 &
  \multicolumn{1}{c|}{Act BPP} &
  \multicolumn{1}{c|}{PSNR} &
  \multicolumn{1}{c|}{MS-SSIM} &
  LPIPS &
  \multicolumn{1}{c|}{Act BPP} &
  \multicolumn{1}{c|}{PSNR} &
  \multicolumn{1}{c|}{MS-SSIM} &
  LPIPS &
  \multicolumn{1}{c|}{Act BPP} &
  \multicolumn{1}{c|}{PSNR} &
  \multicolumn{1}{c|}{MS-SSIM} &
  LPIPS \\ \hline
BPG444 &
  \multicolumn{1}{c|}{0.075} &
  \multicolumn{1}{c|}{27.669} &
  \multicolumn{1}{c|}{0.912} &
  0.363 &
  \multicolumn{1}{c|}{0.15} &
  \multicolumn{1}{c|}{29.711} &
  \multicolumn{1}{c|}{0.945} &
  0.276 &
  \multicolumn{1}{c|}{0.3} &
  \multicolumn{1}{c|}{32.008} &
  \multicolumn{1}{c|}{0.967} &
  0.188 \\ \hline
Cheng2020 &
  \multicolumn{1}{c|}{0.107} &
  \multicolumn{1}{c|}{29.541} &
  \multicolumn{1}{c|}{0.947} &
  0.285 &
  \multicolumn{1}{c|}{0.15} &
  \multicolumn{1}{c|}{30.718} &
  \multicolumn{1}{c|}{0.959} &
  0.241 &
  \multicolumn{1}{c|}{0.299} &
  \multicolumn{1}{c|}{33.261} &
  \multicolumn{1}{c|}{0.978} &
  0.178 \\ \hline
Ours &
  \multicolumn{1}{c|}{0.075} &
  \multicolumn{1}{c|}{27.282} &
  \multicolumn{1}{c|}{0.914} &
  \textbf{0.084} &
  \multicolumn{1}{c|}{0.15} &
  \multicolumn{1}{c|}{28.947} &
  \multicolumn{1}{c|}{0.947} &
  \textbf{0.054} &
  \multicolumn{1}{c|}{0.3} &
  \multicolumn{1}{c|}{31.133} &
  \multicolumn{1}{c|}{0.969} &
  \textbf{0.035} \\ \hline
\end{tabular}%
}
\end{table}
\section{Experiments}
\subsection{Training Settings}
We utilize Flicker2W \cite{liu2020unified}  and select 50,000 images with a size larger than 256 $\times$ 256 from ImageNet \cite{deng2009imagenet} as the training data, totaling approximately 70,000 images. GAN-based image compression task often suffers from unstable training and undesired objective performance, especially at low bitrate. Following HiFiC, we perform a two-stage training strategy to achieve stable training. First, we only use $\mathcal{L}_{rec}$ and $\mathcal{L}_{lpips}$ with a higher $\lambda$ to optimize the compression model 20 epochs. After the compression network warmup, the total loss $\mathcal{L}$ is applied for training the overall network 80 epochs. We adopt Adam optimizer with a batch size of 16 to optimize the network and randomly crop the training sequences into the resolution of 256 × 256. The initial learning rate is set as $1\times10^{-4}$. After 60 epochs, the learning rate is reduced to $1\times10^{-5}$, and the learning rate is reduced to $1\times10^{-6}$ for the last 20 epochs. All experiments are conducted with a single GeForce RTX 3090 GPU.
\begin{figure*}[htbp]
    \centering
    \includegraphics[width=\textwidth]{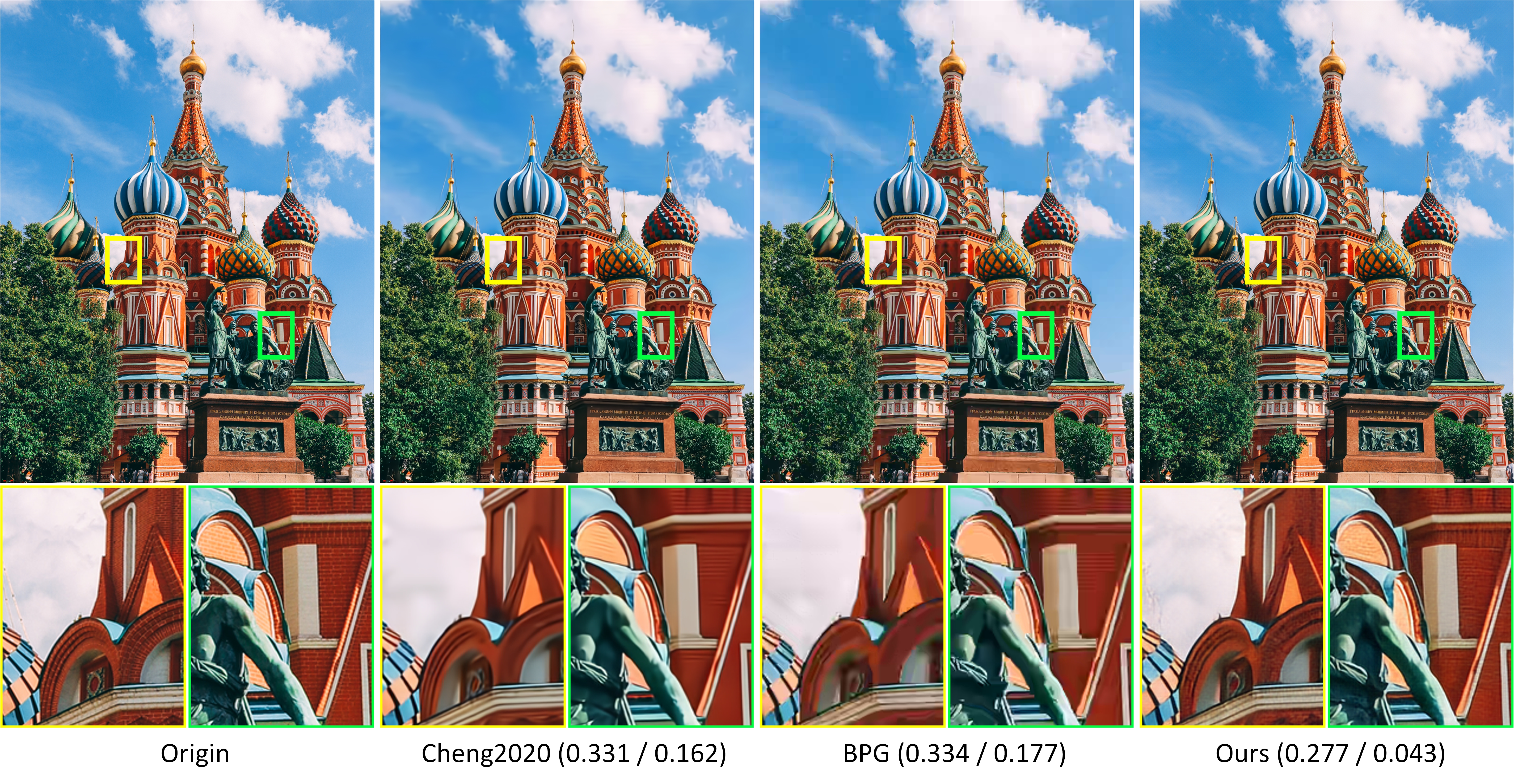}
    \caption{The visual comparison of our method with Cheng2020 and BPG. Our method has more details for the texture of buildings. The parentheses contain (BPP / LPIPS).
    } 
    \label{fig:vis2}
\end{figure*}

\begin{figure*}[htbp]
    \centering
    \includegraphics[width=\textwidth]{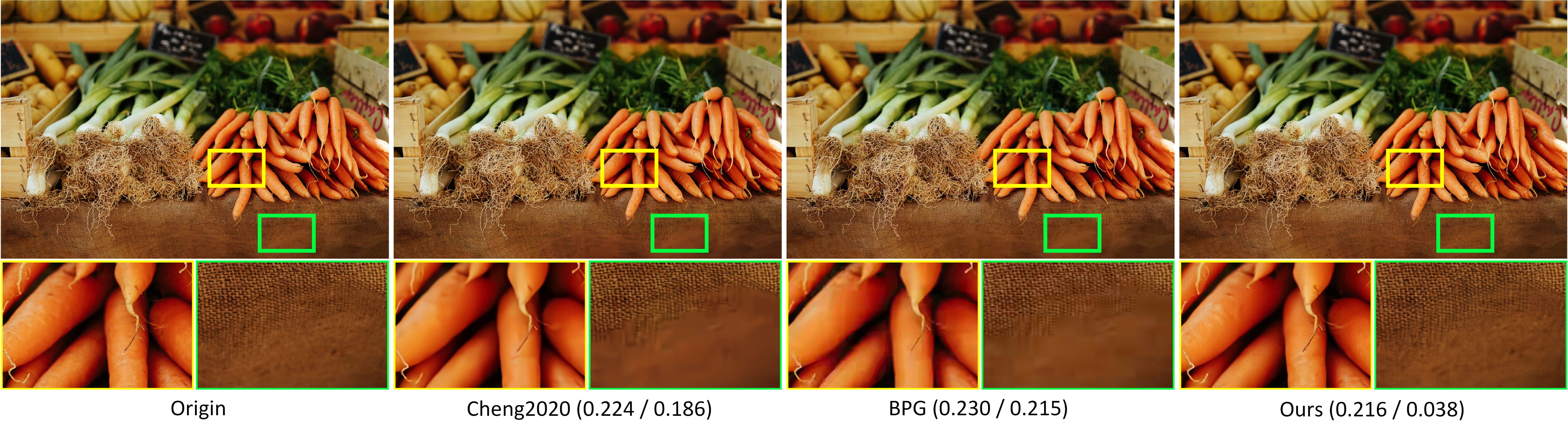}
    \caption{The visual comparison of our method with Cheng2020 and BPG. Our method has more details for the texture of carrots with yellow rectangles and a clearer texture for the linen with green rectangles. The parentheses contain (BPP / LPIPS).
    } 
    \label{fig:vis1}
\end{figure*}
\subsection{Quantitative Results}

We compare our methods with the traditional hybrid compression method BPG and representative learned image compression method Cheng2020 \cite{cheng2020learned}, and all experiments are conducted on CLIC 2024 validation set for the image track. To verify the performance of our proposed schemes, the objective results of different compression methods are shown in Table \ref{tab:obj-results-table}. It can be observed that our method performs the best in terms of LPIPS among all methods.


\subsection{Qualitative Results}
We visually compare our method with BPG and Cheng2020, and experiments demonstrate that our method has higher fidelity at even lower bitrate. Figure \ref{fig:vis1} shows our method has more details for the texture of carrots with yellow rectangles and clearer texture for the linen with green rectangles. Figure \ref{fig:vis2} gives another example.

\section{Conclusions}
In this paper, we propose the perceptual-oriented learned image compression method PO-DKIC. To be specific, for improving perceptual quality and higher fidelity, LPIPS and PatchGAN are utilized to generate more plausible results. PO-DKIC is mainly focusing on perceptual-friendly image compression for human vision. Experiments have shown that our method can achieve superior perceptual quality and obtain high-fidelity images with more texture.

\Section{References}
\bibliographystyle{IEEEbib}
\bibliography{refs}

\end{document}